\pdfoutput=1
\documentclass[11pt]{article}

\usepackage[preprint]{acl}

\usepackage{times}
\usepackage{latexsym}
\usepackage{amsfonts}
\usepackage{amsmath}
\usepackage{microtype}
\usepackage{hyperref}
\usepackage{url}
\usepackage{booktabs}
\usepackage{lineno}
\usepackage{listings}
\usepackage{xcolor}
\usepackage{graphicx}
\usepackage{soul}
\usepackage{enumitem}
\usepackage{setspace}

\usepackage[T1]{fontenc}
\usepackage[utf8]{inputenc}

\usepackage{microtype}

\usepackage{inconsolata}

\usepackage{graphicx}

\title{MHTS: Multi-Hop Tree Structure Framework for Generating Difficulty-Controllable QA Datasets for RAG Evaluation}

\author{
 \textbf{Jeongsoo Lee\textsuperscript{1}*},
 \textbf{Daeyong Kwon\textsuperscript{1,2}*},
 \textbf{Kyohoon Jin\textsuperscript{1}},
 \textbf{Junnyeong Jeong\textsuperscript{1}}
 \textbf{Minwoo Sim\textsuperscript{1}},
 \textbf{Minwoo Kim\textsuperscript{1}},
\\
 \textsuperscript{1}DATUMO,\\ 
 \textsuperscript{2}Graduate School of Culture Technology, Korea Advanced Institute of Science \& Technology,
\\
 \textsuperscript{1}\{jeongsoo.lee, daeyong.kwon, kyohoon.jin, junnyeong.jeong, minwoo.sim, mwkim\}@selectstar.ai
\\
 \small{
   \textbf{Correspondence:} \href{mailto:mwkim@selectstar.ai}{mwkim@selectstar.ai} \& * denotes equal contribution
 }
}

\begin{document}
\maketitle
\begin{abstract}
Existing RAG benchmarks often overlook query difficulty, leading to inflated performance on simpler questions and unreliable evaluations. A robust benchmark dataset must satisfy three key criteria: quality, diversity, and difficulty, which capturing the complexity of reasoning based on hops and the distribution of supporting evidence. In this paper, we propose MHTS (Multi-Hop Tree Structure), a novel dataset synthesis framework that systematically controls multi-hop reasoning complexity by leveraging a multi-hop tree structure to generate logically connected, multi-chunk queries. Our fine-grained difficulty estimation formula exhibits a strong correlation with the overall performance metrics of a RAG system, validating its effectiveness in assessing both retrieval and answer generation capabilities. By ensuring high-quality, diverse, and difficulty-controlled queries, our approach enhances RAG evaluation and benchmarking capabilities.
\end{abstract}

\section{Introduction}
Recent advancements in retrieval-augmented generation (RAG) have significantly enhanced the capabilities of large language models (LLMs) by enabling them to incorporate external knowledge~\citep{brown2020language, lewis2020retrieval, gao2023retrieval, sharma2024retrieval}. These systems are now widely used in open-domain question answering, customer support, and domain-specific retrieval tasks~\citep{karpukhin2020dense, meng2025using, sen2024interactive}. As the adoption of RAG systems continues to grow, evaluating their effectiveness becomes increasingly critical~\citep{izacard2021leveraging}. Existing RAG benchmarks have primarily focused on factual consistency and retrieval relevance, offering metrics that assess whether the generated answers align with the retrieved documents and ground-truth responses~\citep{simon2024methodology, yu2024evaluation}. However, these benchmarks often assume that all queries are of equal difficulty, overlooking the inherent variability in question complexity. These limitations highlight the need for a more holistic evaluation paradigm. In particular, performance differences across question difficulty levels must be examined to properly assess RAG systems.

A robust RAG benchmark must satisfy three core properties: First, it must ensure the \textbf{quality} of the data by providing ground truth (GT) answers that are both accurate and complete. Second, it should promote \textbf{diversity} by covering a wide range of semantic domains, thereby avoiding bias toward narrow topical distributions and enabling more generalizable evaluations. Third, and perhaps most critically, it needs to reflect varying levels of \textbf{difficulty}, capturing the complexity of reasoning required to arrive at correct answers. Despite its importance, the concept of difficulty remains particularly underexplored in recent studies and existing benchmark datasets. Prior research tends to define difficulty using coarse metrics, such as the number of reasoning hops or the count of supporting evidence~\citep{yang2018hotpotqa, tang2024multihop}. However, such research often overlook the semantic complexity and logical integration required to produce a high-fidelity answer.

In practice, answering a complex question often requires synthesizing information scattered across multiple, semantically distant documents~\citep{lu2019answering, de2019question}. The cognitive and computational cost of such synthesis grows not just with the number of reasoning steps but also with the semantic dispersion of the supporting evidence. In particular, reasoning across documents from different topical clusters is typically more demanding than connecting closely related passages. For example, answering a multi-hop question like \textit{“What legal implications has the use of facial recognition technology had in European countries?”} requires synthesizing technical documents on facial recognition systems with legal texts or policy reports from EU jurisdictions. Therefore, a fine-grained measure of query difficulty must take both factors into account: the number of multi-hop reasoning steps and the semantic spread of evidence across source chunks.

To address the lack of fine-grained difficulty control in existing RAG benchmarks, we introduce \textbf{MHTS} (Multi-Hop Tree Structure), a novel framework for synthesizing QA datasets with explicit control over \textbf{query difficulty}. Our approach follows an answer-first generation strategy: we first construct answers requiring multi-hop reasoning across semantically diverse evidence chunks, then generate corresponding queries. This ensures that the reasoning process demands logical integration across dispersed information. We further define a difficulty estimation formula that jointly models two key dimensions of complexity: the number of reasoning hops (i.e., distinct evidence chunks involved) and the semantic distance between the query and each supporting chunk. By capturing both structural and semantic reasoning factors, our difficulty score aligns closely with real RAG performance trends, offering a practical tool for dataset curation and system evaluation.

In addition to enabling fine-grained difficulty control, we validate the quality and diversity of our synthesized dataset. For \textbf{quality}, we ensure that GT answers are logically complete and reflect a comprehensive understanding of the original source documents, providing full answers to the evaluation questions. For \textbf{diversity}, we embed the multi-hop claims that form the backbone of each answer, along with the associated document chunks and their semantically related multi-level summaries, into a shared semantic space. This embedding process reveals that our evaluation dataset (Q, GT) not only spans a wide range of the semantic spectrum but also covers underrepresented regions—thereby ensuring balanced and enriched semantic coverage across the document corpus. While our primary contribution lies in the data generation framework, its implications extend to a broad range of systems, including recent developments in agent-based and modular RAG architectures that require complex multi-hop reasoning. Our benchmark provides a principled foundation for evaluating such systems under varying reasoning difficulty levels.

In summary, this work contributes:
\begin{itemize}
\item \textbf{Fine-grained dataset synthesis framework: } We design a data generation framework that controls reasoning difficulty by combining evidence chunks with varying semantic distances, enabling precise difficulty labeling.

\item \textbf{Validated difficulty metric:} We introduce a new metric capturing both reasoning hops and semantic distance, which aligns well with actual RAG performance trends.

\item \textbf{High-quality benchmark dataset:} Our dataset offers logically complete answers, topic diversity, and multi-hop complexity, making it a robust benchmark for evaluating RAG systems.
\end{itemize}

\section{Related Work}
% FactKG, Multi-Hop RAG, Raptor, HotpotQA, MetaQA

\subsection{Multi-Hop QA Datasets and Generation Methods}

Multi-hop QA requires reasoning across multiple pieces of evidence to answer complex questions, often involving logical composition or information synthesis. Datasets such as HotPotQA~\cite{yang2018hotpotqa}, 2WikiMultiHopQA~\cite{ho2020constructing}, and MuSiQue~\cite{trivedi2022musique} have advanced this area by introducing multi-hop structures and supporting evidence. However, many of these allow shortcut-based solutions or lack structured difficulty control. More recent work like MultiHop-RAG~\cite{tang2024multihop} introduces LLM-generated multi-hop queries with supporting evidence, but still lacks explicit mechanisms to model and control reasoning complexity. Our work addresses this gap by introducing a tree-based framework that enables fine-grained control over multi-hop query difficulty.

\begin{figure*}[t]
\centering
\includegraphics[width=\linewidth]{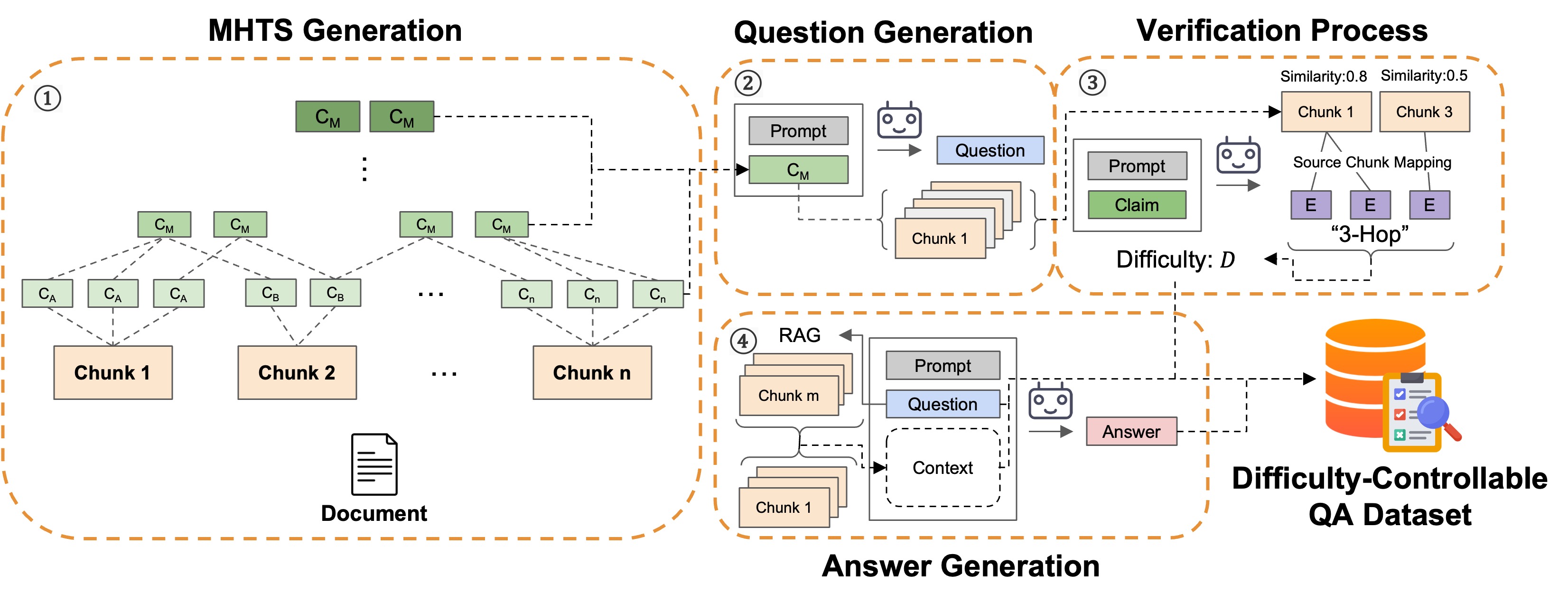}
\caption{Overall process of the Multi-hop Tree Structure (MHTS) framework which synthesizes RAG evaluation dataset by fine-grained difficulty adjustment. \textcircled{1} The source document is chunked and claims are extracted to construct a hierarchical multi-hop tree. \textcircled{2} Each multi-hop claim is used to generate a corresponding question. \textcircled{3} The claim is decomposed into atomic evidences, which are traced back to supporting chunks to estimate reasoning difficulty. \textcircled{4} These verified chunks, combined with top retrieved chunks, are used to synthesize the final answer, forming a complete QA pair.}
\label{fig:main_fig}
\end{figure*}

\subsection{Difficulty Adjustment in Evaluation Datasets}

As QA systems continue to improve, evaluation datasets must evolve to maintain their discriminative power. Prior work has attempted to increase task difficulty through compositional filtering, unanswerable contrast questions~\cite{trivedi2022musique}, or adversarial data collection, where humans iteratively craft questions to fool models~\cite{bartolo2020beat}. 

More recent efforts have explored predicting query difficulty dynamically, such as multHP~\cite{samadi2023performance}, which estimates question complexity prior to retrieval. Though useful for tuning system parameters or balancing query sets, these approaches still do not offer fine-grained, controllable representations of difficulty. Our work addresses this limitation by proposing a structured generation framework that explicitly models and manipulates difficulty during dataset synthesis.

\subsection{Fine-Grained Reasoning Evaluation in RAG Systems}

Retrieval-augmented generation (RAG) systems consist of modular components—retrieval and generation—which require evaluation methods beyond single aggregate scores. To address this, recent studies have proposed diagnostic metrics that separately assess retrieval and answer quality~\cite{ru2025ragchecker}, improving alignment with human judgments and aiding system design. Other work leverages large language models (LLMs) as evaluators. For instance, ARES~\cite{saad2023ares} and related efforts~\cite{liu2025judge} prompt LLMs to assess aspects such as relevance, correctness, and hallucination, offering interpretable, multi-dimensional evaluation. More robust setups, like CONQRET~\cite{dhole2024conqret}, mitigate prompt sensitivity via multi-judge ensembles to enhance consistency across domains. While prior approaches provide detailed post hoc evaluations, they lack control over input complexity. Our framework complements this by explicitly encoding multi-hop reasoning difficulty, enabling more systematic and fine-grained evaluation of RAG systems.

\subsection{Benchmarking RAG systems's Retrieval and Synthesis}

Evaluating retrieval-augmented generation (RAG) system requires not only assessing answer correctness but also how effectively relevant evidence is retrieved. Benchmarks such as KILT~\cite{petroni2020kilt}, MultiHop-RAG~\cite{tang2024multihop}, and CRUD-RAG~\cite{lyu2025crud} have advanced this goal by jointly measuring retrieval and generation performance across a range of knowledge-intensive tasks. These efforts highlight persistent challenges in multi-hop reasoning, long-context understanding, and real-world applicability, with recent metrics like Key Point Recall (KPR)~\cite{qi2024long} aiming to capture more nuanced evidence utilization. Most existing benchmarks lack explicit control over multi-hop difficulty. Our tree-based framework (MHTS) fills this gap by enabling fine-grained difficulty modeling, allowing for more diagnostic evaluation of RAG systems.

\section{Methodology}
%Our method constructs multi-hop QA data through a structured pipeline that recursively performs claim extraction, semantic clustering, and multi-hop composition to build a hierarchical multi-hop structure, from which questions and answers are synthesized based on the resulting multi-hop claims.
Our method constructs multi-hop QA data through a structured pipeline. First, claims are extracted from the source text. Then, semantic clustering and multi-hop composition are recursively performed to build a hierarchical multi-hop structure. Finally, questions and answers are synthesized based on the resulting multi-hop claims. This recursive process enables fine-grained control over reasoning complexity and question difficulty. An overview of the process is illustrated in Figure~\ref{fig:main_fig}.

\subsection{Claim Extraction and Clustering}

To prepare each document for claim-level synthesis, we first segment it into smaller, semantically coherent chunks. This splitting not only reduces the computational overhead for large language models (LLMs) but also ensures that each chunk contains self-contained information. Specifically, we divide the document based on logical boundaries, resulting in a set of chunks $\{C_1, C_2, \dots, C_n\}$.

Next, each chunk $C_i$ is fed into an LLM to extract factual statements (\textit{claims}) spanning five reasoning categories from \cite{kim2023factkg}—namely, \textit{one-hop}, \textit{conjunction}, \textit{existence}, \textit{multi-hop}, and \textit{negation}. During this extraction process, we perform \textit{decontextualization} to minimize ambiguity~\cite{trivedi2022musique}, replacing pronouns or placeholders with explicit expressions. Each extracted claim is mapped back to the index $ix_{C_i}$ of its source chunk $C_i$, enabling us to trace the origin of every claim.

Once the set of claims has been collected, we cluster semantically similar statements to facilitate downstream multi-hop claim generation. Following \citet{sarthi2024raptor}, each claim $c_i$ is embedded into a $d$-dimensional vector $E_i \in \mathbb{R}^d$ using \textit{OpenAI's text-embedding-3-small} model.\footnote{\url{https://platform.openai.com/docs/guides/embeddings}} We then fit a Gaussian Mixture Model (GMM) with $K$ components to these embeddings. Let $\alpha_k$, $\mu_k$, and $\Sigma_k$ respectively denote the mixture weight, mean vector, and covariance matrix for cluster $k$. In this framework, the posterior probability that $c_i$ belongs to cluster $k$ is given by:
\begin{equation}
\begin{aligned}
    \gamma_i(k) 
    \;=\; P\bigl(z_i = k \,\bigm|\; E_i\bigr) \\
    \;=\; \frac{\,\alpha_k\, \mathcal{N}\bigl(E_i \mid \mu_k, \Sigma_k\bigr)\,}
               {\sum_{j=1}^K \alpha_j\, \mathcal{N}\bigl(E_i \mid \mu_j, \Sigma_j\bigr)},
\end{aligned}
\end{equation}
where $z_i$ is the latent cluster variable for claim $c_i$. Since a single claim may be relevant to multiple clusters, we adopt a threshold-based soft assignment: claim $c_i$ is included in cluster $k$ if $\gamma_i(k) \ge \theta$, where $\theta$ is a predefined probability threshold. This design allows overlapping cluster membership and avoids discarding nuanced information. By capturing semantic similarities in this way, we obtain a structured view of how claims interrelate. In subsequent stages, these cluster assignments enable fine-grained multi-hop claim synthesis.

\subsection{Multi-Hop Claim Generation}

Having grouped the extracted claims into clusters, we next generate \emph{multi-hop claims} that synthesize information from multiple source claims. Let $\mathcal{C}^{(k)} = \{c_1, \dots, c_m\}$ be the set of claims assigned to cluster $k$. We provide $\mathcal{C}^{(k)}$ as context to a large language model (LLM) along with guidelines that encourage the model to create statements reflecting multiple reasoning steps. However, we do not impose a strict requirement on the minimum number of source claims to be combined; the LLM is free to decide how many and which claims from $\mathcal{C}^{(k)}$ to incorporate into each new statement.

Formally, let $ \mathcal{G} $ be a generative function parameterized by the LLM. We define a multi-hop claim $c_{\mathrm{mh}}$ as:
\begin{equation}\label{eq:multi_hop_def}
    c_{\mathrm{mh}} = \mathcal{G}\bigl(\{\,c_i \mid i \in S\}\bigr),
\end{equation}
where $S \subseteq \{1,\dots,n\}$ indexes the subset of source claims selected by the LLM. In practice, the model may combine facts, logical inferences, or even negations from multiple $c_i$ to form $c_{\mathrm{mh}}$. During generation, we retain the mapping between each contributing claim $c_i$ and the set of chunks ${C_j}$ from which it was derived. 

After processing all clusters, the newly created multi-hop claims can be integrated back into subsequent iterations if desired. These newly generated statements can then be fed back into the same pipeline—clustering them again to produce yet more complex reasoning required statements. Repeating this procedure yields increasingly sophisticated layers of reasoning, effectively building a \emph{claim-based MHTS (Multi-Hop Tree Structure)}. At each level of MHTS, the LLM is exposed to claims of growing complexity, thereby fostering deeper logical inferences. This iterative approach provides fine-grained control over how many hops of reasoning are embedded in each newly formed statement, ultimately enabling the design of more challenging QA tasks for downstream retrieval-augmented generation systems.

\subsection{Question and Answer Generation}
The generated multi-hop claims capture the core essence of multi-hop reasoning within the document. We therefore treat each claim as a potential “answer,” focusing on the directly connected chunks that contributed to its creation. To construct a QA pair, we first prompt an LLM with the multi-hop claim alone to generate a question that logically leads to it. 

However, because our claim-generation and clustering processes focus on extracting and merging key facts, certain contextual details may be lost. To recover a more comprehensive answer, we perform an \emph{answer decontextualization} procedure. Specifically, for each generated question, we identify the top three chunks most similar to the question (based on a similarity metric) and merge them with the top three chunks associated with the claim that gave rise to the question. If there is overlap between these two sets, the final number of chunks may be fewer than six. This set of chunks is then provided as context to the LLM, which reformulates or expands the claim into a fully contextualized "answer". By grounding the final answer in multiple, highly relevant chunks, we ensure faithfulness to the original source text.

\subsection{Difficulty via Multi-hop}

To confirm that each QA pair truly captures multi-hop reasoning and to quantify its complexity, we decompose the “claim” into a sequence of \emph{evidences}. Following four inference types—one-hop, conjunction, existence, and negation—we exclude ``multi-hop'' itself because it can be represented as a combination of simpler one-hop evidences. Next, we check the entailment relationship between each decomposed evidence and the candidate chunk list. If a piece of evidence cannot be mapped to any chunk, we treat it as hallucination and exclude it through a verification process. However, if the preceding and following evidences map correctly but one intermediate evidence does not, we classify it as an \emph{implicit supposition}; although not explicitly stated in any chunk, it can be inferred from the surrounding context, so we retain it. 

The total \emph{hop count} is determined by the number of these decomposed evidences. By distinguishing between direct mappings, hallucinations, and implicit suppositions, our approach verifies both the factual grounding of each claim and the depth of multi-hop inference required to answer the generated questions. This final check ensures that our QA dataset accurately reflects the intended hierarchical complexity, providing a more reliable benchmark for retrieval-augmented generation systems.

%%%%%%%%%
We defined the \textbf{difficulty} of question-answer pairs using two variables. The first variable is the number of pieces of evidence used in the answer. To determine this, we decomposed the multi-hop claims containing the core information of the answer into pieces of evidence based on the four reasoning types in \citet{kim2023factkg}: one-hop, conjunction, existence, and negation. The number of decomposed evidence pieces is referred to as the "number of hops."

The second variable is the similarity between the question and the supporting chunks used in the answer. To compute this, we first retrieved supporting chunks for each piece of decomposed evidence. Then, we calculated the cosine similarity between the question embedding and each supporting chunk and took the average. If a single piece of evidence was supported by multiple chunks, we used the average similarity across those chunks.

The final difficulty score $D$ is defined as $D = h - \lambda s$ where $h$ is the number of hops, $s$ is the average similarity, $\lambda$ is a scaling factor (set to 1 by default).

\section{Experiments}
% 우리는 생성된 QA 데이터셋을 Quality, Difficulty, Diversity 측면에서 평가합니다.

We evaluate the proposed MHTS dataset along three dimensions:  (1) \textbf{Difficulty Calibration}, by demonstrating how our difficulty scores accurately reflect reasoning complexity through fine-grained analysis of win-rates (GT vs. RAG) across varying hop counts;  (2) \textbf{Semantic Diversity}, by quantifying how comprehensively the generated dataset spans the underlying knowledge space; and (3) \textbf{Answer Quality}, through qualitative analysis illustrating that our method captures complex multi-hop reasoning and inference capabilities absent in RAG outputs.

\subsection{Experimental Setup}

\subsubsection{Dataset}

The novel \textit{David Copperfield}~\footnote{\url{https://www.gutenberg.org/ebooks/766}} has been chosen as a dataset for QA generation. David Copperfield provides various character relationships and distributed event information, making it highly suitable for creating a multi-hop QA dataset. The complex narrative structure of the novel and the interconnectedness of the information provide an ideal environment for tasks that require multi-hop reasoning, where the QA system must go through several steps to derive an answer. The entire data was divided by paragraph, and then concatenated to form chunks with a maximum length of 1024 tokens. As a result, 505 chunks were created, with a minimum length of 540 tokens and a maximum length of 1016 tokens.

\subsection{Baseline and Ground-Truth (GT) Context}

\paragraph{RAG Setup}
Our baseline employs a two-stage retrieval pipeline. Initially, we retrieve the top-10 chunks by ranking their cosine similarity scores against the input question~\citep{lewis2020retrieval}. Subsequently, a reranker~\footnote{\url{https://huggingface.co/jinaai/jina-reranker-v2-base-multilingual}} selects the three most relevant chunks to form the final context~\citep{glass2022re2g}.

\paragraph{Ground-Truth (GT) Context}
Our ground-truth (GT) context leverages gold evidence chunks, mapped explicitly during claim construction. To ensure optimal relevance, we select the three chunks most similar to the input question among all mapped evidence chunks. These selected gold chunks typically capture key reasoning steps essential for accurate multi-hop inference—steps that naive embedding-based retrieval methods frequently miss. To further enrich the GT context and allow comparative evaluation, we include an additional three chunks retrieved by the RAG pipeline. 
%Consequently, the final GT context comprises six chunks, enabling a comprehensive assessment of retrieval effectiveness and highlighting the advantages of our structured, tree-based chunk mapping approach over surface-level similarity retrieval.
Consequently, the final GT context consists of up to six chunks. However, if there is overlap between the retrieved chunks and those already associated with the claim, the total number may be fewer than six. This ensures a comprehensive assessment of retrieval effectiveness while highlighting the advantages of our structured, tree-based chunk mapping approach over surface-level similarity retrieval.

% \subsubsection{Difficulty \& Quality}

% To demonstrate the superiority of our generated answers, we used a naive RAG approach to generate answers by retrieving the top 3 most similar chunks based on their similarity to the question and providing them as context. We then asked GPT-4o to evaluate the question, our answer, and the naive RAG answer, selecting the winner for three metrics to measure the win rate.

% The three metrics—\textit{Comprehensiveness, Diversity, and Empowerment}—were adopted from ~\cite{edge2024local}. To mitigate positional bias, we swapped the order of our answer and the naive RAG answer and repeated the evaluation twice. The final classification into win/tie/lose was determined by aggregating the results from both evaluations. To verify that higher hop counts correspond to more challenging QA tasks in multi-hop QA, we compared the changes in win rate based on the hop count.

\subsection{Fine-grained Difficulty}
% MHTS-Raptor와 MHTS-Claim의 QA를 설정한 metric에 대해서 LLM evaluator를 통해 평가하고, 평균 점수를 계산합니다. Metric은 포괄성, 구체성, 유창성, 사실성 등이 후보입니다. 

% Comprehensiveness: How much detail does the answer provide to cover all aspects and details of the question?
% Specificity (구체성): 정의: 답변이 얼마나 구체적인지를 평가하는 지표입니다. 질문에 대한 답변이 일반적인 답변을 넘어, 특정한 상황이나 세부 사항을 다루고 있는지를 확인합니다. 중요성: 구체적인 답변은 질문에 대한 명확하고 정확한 정보를 제공하며, 불필요하게 모호하거나 지나치게 일반적인 답변을 피할 수 있습니다.
% Fluency (유창성): 정의: 답변의 언어가 얼마나 자연스럽고 유창한지를 평가하는 지표입니다. 문장 구조나 단어 선택이 어색하지 않고 자연스러운 흐름을 이루는지를 판단합니다. 중요성: 유창한 답변은 읽기 쉽고 이해하기 쉬워집니다. 어색한 표현이나 부자연스러운 문장은 독자가 답변을 이해하는 데 어려움을 겪을 수 있습니다.
% Factuality (사실성): 원본 청크와 QA를 넣고 답변의 내용에 hallucination이 있는지 체크

To verify that our method generates questions in a fine-grained manner according to varying difficulty levels, and simultaneously ensures that the generated answers satisfy high-quality standards, we employed an LLM-as-a-Judge evaluation approach~\citep{zheng2023judging}. Specifically, we presented GPT-4o~\footnote{\url{https://openai.com/index/hello-gpt-4o/}} with a question, our proposed answer, and a RAG answer, prompting it to select the better response according to three metrics, thereby measuring the win rate.

The three metrics—\textit{Comprehensiveness}, \textit{Diversity}, and \textit{Empowerment}—were adopted from~\cite{edge2024local}. To mitigate potential positional bias, we repeated the evaluation twice, swapping the order in which our answer and the naive RAG answer were presented. The final classification into ``win/tie/lose" was determined by aggregating results from both evaluations. Additionally, to confirm that higher hop counts correspond to increased difficulty in multi-hop QA tasks, we analyzed how the win rate varied across different hop counts.

\begin{figure}[t]
\centering
\includegraphics[width=0.99\columnwidth]{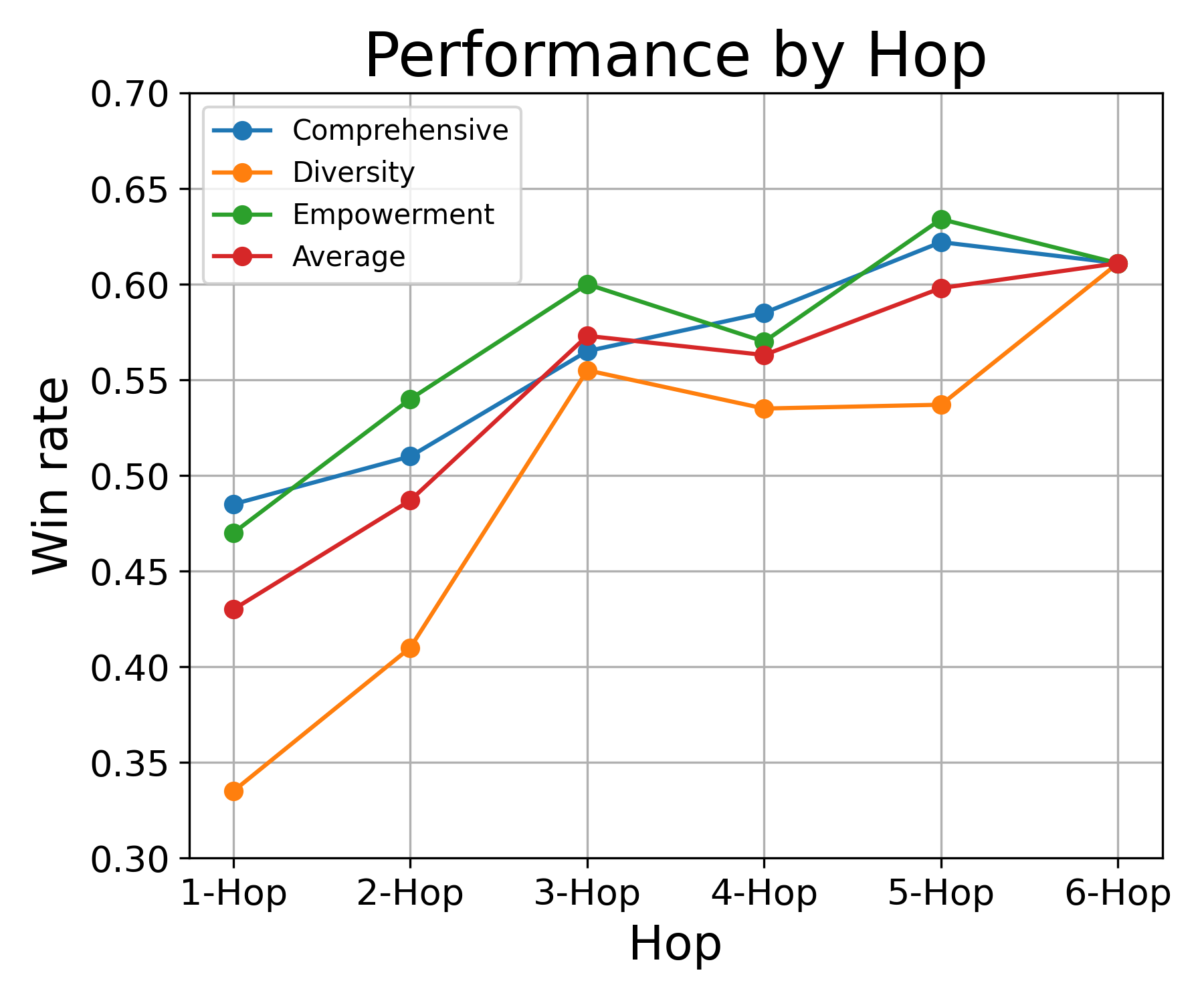}
\caption{GT win rate increases with hop count (Pearson \(r = 0.93\)).}
\label{fig:hop_win_plot}
\end{figure}

\begin{figure}[t]
\centering
\includegraphics[width=0.99\columnwidth]{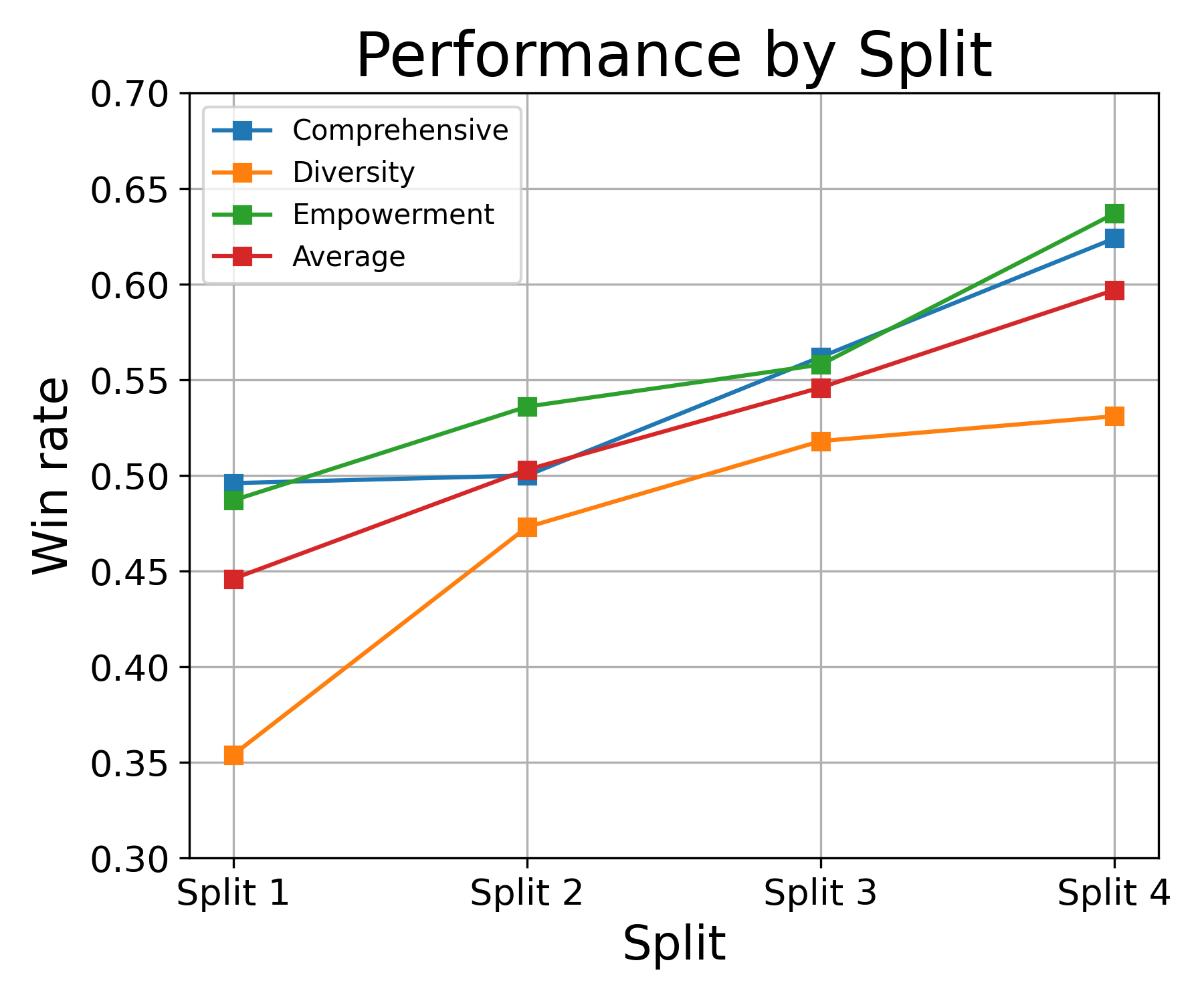}
\caption{GT win rate increases with difficulty (Pearson \(r = 0.99\)).}
\label{fig:hop_win_plot}
\end{figure}

Figure~\ref{fig:hop_win_plot} illustrates that our generated (GT) answers consistently outperform naive RAG outputs across various evaluation metrics—\textit{Comprehensiveness}, \textit{Diversity}, and \textit{Empowerment}. Importantly, we observe a clear positive correlation between the difficulty level and the GT win rate. This trend highlights that as the complexity of multi-hop reasoning tasks increases, naive RAG systems struggle more significantly, either failing to provide adequate answers or delivering incomplete responses. In contrast, our MHTS-generated answers not only remain robust but become increasingly advantageous at higher difficulty levels.

However, when evaluating the win rate strictly based on hop count (Figure~\ref{fig:hop_win_plot} (a)), although an overall increasing trend is observable, there are fluctuations where the win rate temporarily decreases before rising again. While the general trend still demonstrates a linear increase, these intermediate drops indicate limitations in relying solely on hop count for fine-grained difficulty control. In contrast, our proposed method (Figure~\ref{fig:hop_win_plot} (b)) exhibits a consistent, linear improvement without fluctuations, achieving a correlation of 0.99 between difficulty level and win rate. Thus, compared to a simplistic hop-count-based approach, our method provides a more reliable and fine-grained difficulty calibration for multi-hop QA. 

In Figure~\ref{fig:hop_win_plot} (a), the evaluation was conducted using 100 samples for hop counts 1 to 4, 41 samples for 5-hop, and 9 samples for 6-hop. For the split-based analysis: Figure ~\ref{fig:hop_win_plot} (b), we selected 90 samples per difficulty level, ensuring an equal number of samples across different difficulty splits.

% \begin{table}[h]
% \centering
% \small
% \begin{tabular}{ccccc}
% \toprule
% \textbf{Hops} & \textbf{Compre.} & \textbf{Diversity} & \textbf{Empower.} & \textbf{Avg.} \\
% \midrule
% 1 & 48.5 & 33.5 & 47.0 & 43.0 \\
% 2 & 51.0 & 41.0 & 54.0 & 48.7 \\
% 3 & 56.5 & 55.5 & 60.0 & 57.3 \\
% 4 & 58.5 & 53.5 & 57.0 & 56.3 \\
% 5 & 62.2 & 53.7 & 63.4 & 59.8 \\
% 6 & 61.1 & 61.1 & 61.1 & 61.1 \\

% \bottomrule
% \end{tabular}
% \caption{GT win rate by hop count (\%).}
% \label{tab:llmjudge_winrate}
% \end{table}

\subsection{Diversity}
% 원천 문서(청크)의 분포와, 생성된 QA의 분포를 비교합니다.
% 클러스터링으로 똑같이 하긴 하지만
% -> 전체적인 논리 구조를 더 잘 보여주고 있다.

Evaluating semantic diversity is critical for understanding the semantic coverage and reasoning breadth and comprehensiveness of a QA dataset. Traditional hierarchical summarization methods such as RAPTOR~\cite{sarthi2024raptor} rely on recursively clustering summarized nodes, which may limit semantic coverage by failing to capture the logical integration across multiple, semantically distinct chunks. To verify whether our MHTS-based method improves upon these limitations by capturing a wider semantic range, we conduct a comparative evaluation against a RAPTOR-style hierarchy.

Specifically, we construct the RAPTOR-style hierarchical structure using 100-token chunks and their summaries, recursively clustering until four root nodes remain. We embed both RAPTOR nodes and our MHTS-generated (Q, A) pairs using the same embedding model for fair comparison, then assess semantic diversity via average pairwise distance (higher indicates broader topic coverage) and the sum of the eigenvalues of the covariance matrix (higher denotes richer semantic coverage).

\begin{table}[t]
\centering
\small
\begin{tabular}{lcc}
\toprule
\textbf{Dataset} & \textbf{Avg. Dis.} ↑ & \textbf{Eigen Var.} ↑ \\
\midrule
RAPTOR Tree Nodes & 1.12 & 0.63 \\
MHTS (Q, A) Pairs & \textbf{1.23} & \textbf{0.76} \\
\bottomrule
\end{tabular}
\caption{Semantic diversity: higher distance and higher variance indicate broader coverage.}
\label{tab:diversity}
\end{table}
\noindent

Table~\ref{tab:diversity} presents the semantic diversity evaluation results. For a fair comparison, the same number of data points were used for both datasets. MHTS-generated (Q, A) pairs exhibit a higher average embedding distance (1.23) compared to RAPTOR nodes (1.12), indicating that our data points are more semantically dispersed and less redundant. In addition, the total sum of eigenvalues is also greater for MHTS (0.76 vs. 0.63), suggesting that the semantic embeddings occupy a richer and more varied subspace.

These results strongly indicate that our MHTS framework effectively generates QA pairs that span a more diverse and comprehensive semantic space than traditional summarization-based hierarchical structures. This broader coverage not only enriches the dataset but also enables more fine-grained evaluation of a model’s ability to perform multi-hop reasoning, ultimately allowing us to assess how well systems handle complex, real-world QA tasks in knowledge-intensive domains. A visual analysis of this semantic diversity is provided in Appendix~\ref{appendix_diversity}, further illustrating how MHTS QA pairs span a broader range of topics compared to RAPTOR nodes.

% 그림을 통한 자세한 분석 내용은 appendix에 추가하겠다 는 내용을 추가하고 그림도 appendix로

\subsection{Quality}

While difficulty calibration and semantic diversity measure important aspects of a QA dataset, high-quality reference answers(GT) are ultimately essential for evaluating system performance in realistic scenarios. To assess whether our MHTS-generated QA pairs maintain strong GT answer quality, we conducted a qualitative comparison against naive RAG outputs.

\begin{table*}[t]
\resizebox{\textwidth}{!}{%
\begin{tabular}{c|c}
\hline
Question   & \begin{tabular}[c]{@{}c@{}}What was the relationship between Emily and the person \\ she was with when the narrator saw her on the moving ship?\end{tabular}                                                                                                                                                                                             \\ \hline
RAG  & \begin{tabular}[c]{@{}c@{}}Based on the provided context, Emily was romantically involved with \textbf{Mr. James}. \\ When the narrator saw her on the moving ship, \\ she was with Mr. James, her former partner. Their relationship had become strained \\ because Mr. James had left Emily to marry another respectable person, \\ as arranged by Mr. Littimer. This departure led to Emily's distress \\ and subsequent actions, highlighting the complexities and conflicts in their relationship.\end{tabular}                                                                                                            \\ \hline
GT (MHTS) & \begin{tabular}[c]{@{}c@{}}Emily was accompanied by \textbf{her uncle} when the narrator saw her on the moving ship. \\ The context describes Emily "trembling on his shoulder" and clinging to him \\ "with the utmost trust of thy bruised heart," indicating a close \\ and trusting relationship between them. \\ The uncle is portrayed as someone who has shown "all the might of his great love" for Emily, \\ highlighting his protective and caring role in her life.\end{tabular}                                                                                         \\ \hline \hline
Chunk      & \begin{tabular}[c]{@{}c@{}}As the sails rose to the wind, and the ship began to move, \\ there broke from all the boats three resounding cheers, which those on board took up, \\ and echoed back, and which were echoed and re-echoed. \\ My heart burst out when I heard the sound, \\ and beheld the waving of the hats and handkerchiefs--and then I saw her! \\ Then I saw her, at \textbf{her uncle}’s side, and trembling on his shoulder. \\ He pointed to us with an eager hand; and she saw us, \\ and waved her last good-bye to me. Aye, Emily, beautiful and drooping, \\ cling to him with the utmost trust of thy bruised heart; \\ for he has clung to thee, with all the might of his great love!\end{tabular} \\ \hline
\end{tabular}%
}
\caption{Case study for quality evaluation. The question is answered using both RAG (\textit{RAG}) and our proposed method based on MHTS (\textit{GT (MHTS)}). The supporting evidence for the question is shown in the bottom row (\textit{Chunk}) which is a mapped chunk for GT. Notably, the relevant chunk was not included in the top-3 retrieved chunks used by naive RAG, leading to an incorrect, hallucinated response.}
\label{tab:quality}
\end{table*}

Table~\ref{tab:quality} presents an illustrative example comparing a Ground Truth (GT) answer to that of a naive RAG system. The question asks about \textit{"the relationship between Emily and the person she was with when the narrator saw her on the moving ship."} However, the top-3 chunks retrieved by naive RAG omit the relevant portion of the text, resulting in a hallucinated answer that describes Emily as being romantically involved with \textit{Mr. James}. In reality, the source text reveals that she was with \textit{her uncle}, portrayed as a deeply trusted and protective figure. The GT answer accurately reflects this relationship, grounded in the original evidence, whereas the RAG response fabricates a misleading narrative due to missing context.

This discrepancy highlights the significance of proper context retrieval and multi-hop integration for answer quality. By leveraging a more comprehensive evidence-mapping strategy, our MHTS-based pipeline avoids the crucial omissions that lead to RAG’s hallucinations, thereby producing answers that align with the factual details of the original source text. Additional examples used for quality evaluation can be found in Appendix~\ref{appendix_quality}.

% 5계층에 따른 청크 수, 홉 수 데이터

% \begin{table}[]
% \centering
% \begin{tabular}{@{}lccccc@{}}
% \toprule
% \textbf{}           & \textbf{1} & \textbf{2} & \textbf{3} & \textbf{4} & \textbf{5} \\ \midrule
% \textbf{\# Hops}    & 0.76              & 1.57       & 2.57       & 3.14       & 3.57              \\
% \textbf{\# Chunks}  & 1.00                 & 1.38       & 1.88       & 2.32       & 2.67              \\
% \textbf{Difficulty} & 0.76              & 2.17       & 4.83       & 7.28       & 9.53              \\ \bottomrule
% \end{tabular}
% \caption{Number of Hops and Chunks for different levels. Level 1 is leaf and level 5 is root.}
% \end{table}

\section{Conclusion}
In this work, we introduced a novel framework—MHTS—for generating multi-hop QA datasets that systematically controls difficulty and ensures both semantic diversity and high-quality answers. By splitting source documents into chunks, extracting and clustering claims, and iteratively generating multi-hop statements, our method builds a tree structure that reflects progressively deeper reasoning. We demonstrated that relying solely on naive retrieval-augmented generation (RAG) methods leads to suboptimal performance when handling complex queries. Specifically, we observed that as the difficulty level—defined by our formulation combining reasoning hops and semantic dispersion—increases, MHTS-based answers achieve higher win rates. These findings highlight the importance of fine-grained difficulty calibration in the development and evaluation of advanced RAG systems, serving as a foundation for more robust and comprehensive research in multi-hop question answering. Since current RAG systems tend to perform poorly and frequently produce hallucinations when faced with high-difficulty queries, the ability to systematically synthesize such challenging examples with fine-grained difficulty control makes our framework particularly valuable for advancing robust and realistic RAG evaluation.

% \section*{Limitation}
% Lorem Ipsum is simply dummy text of the printing and typesetting industry. Lorem Ipsum has been the industry's standard dummy text ever since the 1500s, when an unknown printer took a galley of type and scrambled it to make a type specimen book. It has survived not only five centuries, but also the leap into electronic typesetting, remaining essentially unchanged. It was popularised in the 1960s with the release of Letraset sheets containing Lorem Ipsum passages, and more recently with desktop publishing software like Aldus PageMaker including versions of Lorem Ipsum.

% \section*{Author Contributions}
% If you'd like to, you may include  a section for author contributions as is done
% in many journals. This is optional and at the discretion of the authors.

% \section*{Acknowledgments}
% Use unnumbered first level headings for the acknowledgments. All
% acknowledgments, including those to funding agencies, go at the end of the paper.

% \section*{Ethics Statement}
% Authors can add an optional ethics statement to the paper. 
% For papers that touch on ethical issues, this section will be evaluated as part of the review process. The ethics statement should come at the end of the paper. It does not count toward the page limit, but should not be more than 1 page.

\section*{Limitation}
While our proposed MHTS framework offers a systematic approach to generating difficulty-controlled QA datasets, it has certain limitations.

First, the current evaluation is conducted exclusively on a single literary corpus—David Copperfield. Although this source provides a rich setting for testing multi-hop reasoning due to its complex narrative structure and character interactions, it may not fully represent the challenges found in other domains such as scientific articles, encyclopedic entries, or real-world dialogues. Future work should explore the applicability of MHTS across a broader range of document types to assess its generalizability.

Second, the framework relies on the performance of large language models (LLMs) for critical components such as claim extraction, clustering, and multi-hop synthesis. While strong results were achieved using powerful LLMs, the quality and controllability of the generated datasets may degrade when using smaller or less capable models. This dependency raises concerns about the framework’s accessibility and robustness in resource-constrained environments.

% Bibliography entries for the entire Anthology, followed by custom entries
%\bibliography{anthology,custom}
% Custom bibliography entries only
% \bibliography{latex/acl_latex}

\appendix

\section{Example Appendix}
\label{sec:appendix}

% You may include other additional sections here.

\begin{figure}[h] 
\centering 
\includegraphics[width=\columnwidth]{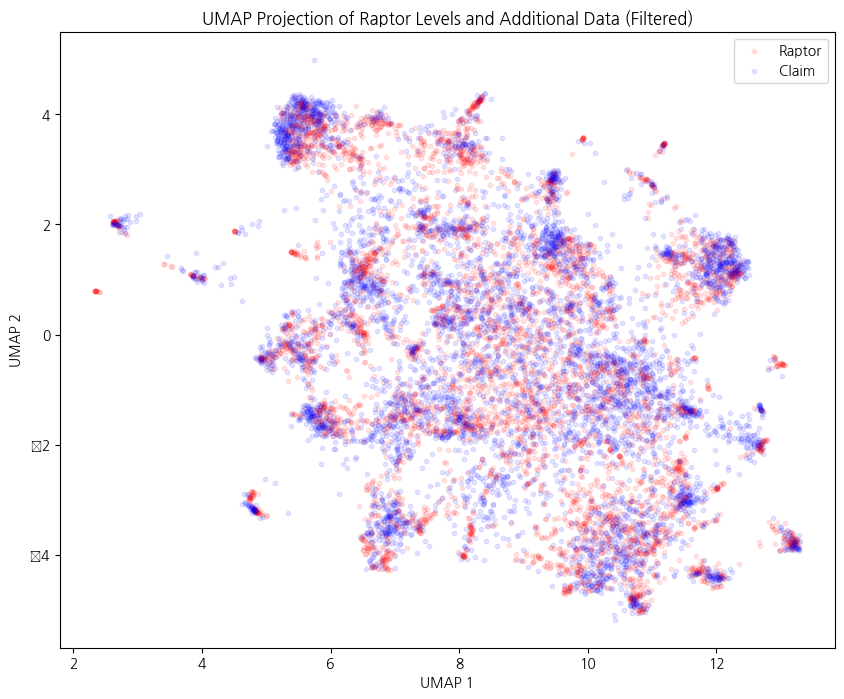} 
\caption{UMAP projection of semantic embeddings from MHTS claims (blue) and RAPTOR nodes (red). Claims appear more semantically dispersed, suggesting broader topic coverage.} \label{fig:umap_diversity} 
\end{figure}

\subsection{Diversity Visualization}
\label{appendix_diversity}

To provide a visual illustration of the semantic diversity of our generated QA data, we project the embeddings of claim-level representations into a 2D space using UMAP. From our full claim set, we sample 5,527 claims via K-means clustering to ensure broad coverage and compare them against node representations extracted from a RAPTOR-style summarization hierarchy~\cite{sarthi2024raptor}. As shown in Figure~\ref{fig:umap_diversity}, although both distributions overlap considerably, the MHTS-generated claims (in blue) appear more dispersed across the semantic space than the RAPTOR nodes (in red), which tend to form tighter clusters. This dispersion suggests a slightly higher degree of topic-level diversity in our data. However, the difference is subtle, and the overall semantic coverage of both datasets remains comparable.

\subsection{Additional Quality Experiments}
\label{appendix_quality}

\begin{table*}[t]
\resizebox{\textwidth}{!}{%
\begin{tabular}{c|c}
\hline
\textbf{Question}   & \begin{tabular}[c]{@{}c@{}}After having breakfast on the Heath, \\ to which location did the narrator walk back and loiter \\ for half an hour before old Tiffey appeared with his key?\end{tabular}                                                                                                                                                                                                                                                                                                                                                                                                                                        \\ \hline 
\textbf{Naiive RAG} & \begin{tabular}[c]{@{}c@{}}After having breakfast on the Heath, \\ the narrator walked back to Doctors’ Commons. \\ Upon arriving at the office there, he loitered around \\ the Commons for half an hour before old Tiffey appeared with his key.\end{tabular}                                                                                                     \\ \hline
\textbf{GT (MHTS)}  & \begin{tabular}[c]{@{}c@{}}After having breakfast on the shore, \\ the narrator walked back and loitered \\ for half an hour before old Tiffey appeared with his key.\end{tabular} \\ \hline \hline
\textbf{Chunk}      & \begin{tabular}[c]{@{}c@{}}I got some breakfast on the Heath, and walked back to Doctors’ Commons, \\ along the watered roads and through a pleasant smell of summer flowers, \\ growing in gardens and carried into town on hucksters’ heads, \\ intent on this first effort to meet our altered circumstances. \\ I arrived at the office so soon, after all, \\ that I had half an hour’s loitering about the Commons, \\ before old Tiffey, who was always first, appeared with his key.\end{tabular}                  \\ \hline
\end{tabular}%
}
\caption{Case study for quality evaluation. The question is answered using both naive RAG (\textit{Naive RAG}) and our proposed method based on MHTS (\textit{GT (MHTS)}). The supporting evidence for the question is shown in the bottom row (\textit{Chunk}). Notably, the relevant chunk was not included in the top-3 retrieved chunks used by naive RAG, leading to an incorrect, hallucinated response.}
\label{tab:quality_appendix1}
\end{table*}

\begin{table*}[h]
\resizebox{\textwidth}{!}{%
\begin{tabular}{c|c}
\hline
\textbf{Question}   & \begin{tabular}[c]{@{}c@{}}What did the nurse and some sage women declare \\ about the consequences of being born on a Friday night for the narrator?\end{tabular}                                                                                                                                                                                                                                                                                                                                                                                                                                        \\ \hline 
\textbf{Naiive RAG} & \begin{tabular}[c]{@{}c@{}}Based on the provided context, Miss Betsey was the one who indicated \\ that being born on a Friday had significant importance for the narrator. \\ She expressed that the narrator owes his birth on a Friday to her actions, \\ suggesting that this timing had positive consequences for his future. \\ There is no mention of a nurse or other sage women declaring anything \\ about the consequences of being born on a Friday night in the given text.\end{tabular}                                                                                                     \\ \hline
\textbf{GT (MHTS)}  & \begin{tabular}[c]{@{}c@{}}The nurse and several wise women in the narrator's neighborhood declared \\ two specific consequences of his birth on a Friday night at midnight. \\ They believed that:\\ 1. **Destined for Unluckiness**: \\The narrator was fated to experience misfortune throughout his life.\\ 2. **Ability to See Ghosts and Spirits**: \\ He was granted the unique privilege of being able to perceive ghosts and spirits.\\ They held that these attributes were inevitably linked to all infants born \\ during the early hours of a Friday night, regardless of gender.\end{tabular} \\ \hline \hline
\textbf{Chunk}      & \begin{tabular}[c]{@{}c@{}}In consideration of the day and hour of my birth, it was declared by the nurse, \\ and by some sage women in the neighbourhood \\ who had taken a lively interest in me several months \\ before there was any possibility of our becoming personally acquainted, \\ first, that I was destined to be unlucky in life; \\ and secondly, that I was privileged to see ghosts and spirits; \\ both these gifts inevitably attaching, as they believed, \\ to all unlucky infants of either gender, \\ born towards the small hours on a Friday night.\end{tabular}                  \\ \hline
\end{tabular}%
}
\caption{Case study for quality evaluation. The question is answered using both naive RAG (\textit{Naive RAG}) and our proposed method based on MHTS (\textit{GT (MHTS)}). The supporting evidence for the question is shown in the bottom row (\textit{Chunk}).}
\label{tab:quality_appendix2}
\end{table*}

Table~\ref{tab:quality_appendix1} presents an illustrative example comparing a Ground Truth (GT) answer to that of a naive RAG system. The question asks about \textit{"the location the narrator returned to and loitered at for half an hour before old Tiffey appeared with his key, after having breakfast on the Heath."} However, the top-3 chunks retrieved by naive RAG omit the relevant portion of the text, resulting in an answer that simply repeats the question’s phrasing and lacks supporting detail. In contrast, the GT answer produced via MHTS correctly identifies that \textit{"the narrator walked back to Doctors’ Commons and loitered there for half an hour,"} as described in the source chunk. 

Table~\ref{tab:quality_appendix2} presents an illustrative example comparing a Ground Truth (GT) answer to that of a naive RAG system. The question asks about the consequences declared by a nurse and some sage women for an infant born on a Friday night. However, the top-3 chunks retrieved by naive RAG omit the relevant portion of the text, resulting in a hallucinated mention of “Miss Betsey” and a falsely optimistic outlook. In contrast, the GT answer correctly identifies that the nurse and the wise women believed the child would be both unlucky and able to see ghosts and spirits.

This discrepancy highlights the significance of proper context retrieval and multi-hop integration for answer quality. By leveraging a more comprehensive evidence-mapping strategy, our MHTS-based pipeline avoids the crucial omissions that lead to RAG’s hallucinations, thereby producing answers that align with the factual details of the original source text.

\end{document}